\documentclass[floatfix,showpacs,amsmath,amsfonts,amssymb,aps,twocolumn,superscriptaddress,prl,10pt]{revtex4-1}
\usepackage[pdftex]{graphicx}
\usepackage{array}

\newcommand{\figref}[1]{Fig.~\ref{#1}}

\newcommand{\diffd}{\text{d}}
\renewcommand{\vec}[1]{\mathbf{#1}}
\newcommand{\punc}[1]{\,#1}
\newcommand{\neweqnline}{\nonumber\\}

\newcommand{\nud}[2]{$\ket{#1,\!#2}$}
\newcommand{\eud}[3]{$\ket{#1,\!#2}^{#3}$}
\newcommand{\vac}{|0\rangle}

\newcommand{\ket}[1]{|{#1}\rangle}
\newcommand{\cdu}[1]{c^{\dagger}_{#1{\uparrow}}}
\newcommand{\cdd}[1]{c^{\dagger}_{#1{\downarrow}}}

\begin{document}

\title{Inhomogeneous state of few-fermion superfluids}
\author{P.O.~Bugnion}
\affiliation{Cavendish Laboratory, J.J. Thomson Avenue, Cambridge, CB3 0HE, United Kingdom}
\author{J.A.~Lofthouse}
\affiliation{Cavendish Laboratory, J.J. Thomson Avenue, Cambridge, CB3 0HE, United Kingdom}
\author{G.J.~Conduit}
\affiliation{Cavendish Laboratory, J.J. Thomson Avenue, Cambridge, CB3 0HE, United Kingdom}
\date{\today}

\begin{abstract} The few-fermion atomic gas is an ideal setting to explore
  inhomogeneous superfluid pairing analogous to the Larkin-Ovchinnikov
  state. Two up and one down-spin atom is the minimal configuration that
  displays an inhomogeneous pairing density whereas imbalanced systems
  containing more fermions present a more complex pairing topology.  With
  more than eight atoms trapped the system approaches the macroscopic
  superfluid limit. An oblate trap with a central barrier offers a direct
  experimental probe of pairing inhomogeneity.
\end{abstract}


\maketitle

The rapidly advancing field of ultracold atomic gases has opened new vistas
of experimentally accessible phases of matter. The observation of
superfluidity~\cite{Greiner03,Jochim03,Zwierlein03,Bourdel04} and density
imbalance~\cite{Zwierlein06,Partridge06} in a two-component Fermi gas
presents the building blocks required to realize the inhomogeneous
superfluid phase proposed by Fulde, Ferrel, Larkin, and Ovchinnikov
(FFLO)~\cite{Fulde64,Larkin65}. This state plays a central role in our
understanding of superconductors, superfluids, and particle
physics~\cite{Casalbuoni04} but has never been unambiguously realized in
solid state superconductors~\cite{Matsuda07}. Theory predicts that the FFLO
phase should be present in a three-dimensional atomic gas, however it has
not been observed~\cite{Zwierlein06,Partridge06}. In a one-dimensional
imbalanced atomic gas, although the superfluid is predicted to be
stable~\cite{Orso07,Hu07}, any inhomogeneous pairing present was too weak to
be observed~\cite{Liao10,Casula08}.  We now exploit the new experimental
capability to trap, manipulate, and address a few fermionic
atoms~\cite{Cheinet08,Serwane11} to propose a protocol to create an
inhomogeneous superfluid. The pairing has a simple nodal structure that can
be directly characterized by experiment.

To study the few-fermion superfluid we take advantage of recent experimental
developments that allow investigators to confine up to ten atoms in a trap
and address their quantum state~\cite{Cheinet08,Serwane11,Zurn12}.  This
presents a unique opportunity to study the microscopic physics of contact
interactions~\cite{Busch98,Mora04,Idziaszek06,Liu10,Rubeni12,
  Rontani12,Gharashi12,Brouzos12} in a tractable setting, and then scale the
intuition up to a many-body system. Following this program, experimentalists
could realize an analog to the Stoner model for itinerant
ferromagnetism~\cite{Stoner38,Bugnion13ii}, and the direct and double
exchange mechanisms~\cite{Bugnion13}. We now turn from repulsive to
attractive interactions to study a counterpart to the BCS
superconductor. Though the three-fermion ground state has been previously
examined~\cite{Stetcu07,Kestner07} in both three and one
dimension~\cite{Baur10}, the authors overlooked the underlying inhomogeneous
pairing.

To orient the discussion we first expose the symmetry changes in the ground
state that foretell the emergence of inhomogeneous pairing. Next we study
the inhomogeneous pairing for a state with $N_{\uparrow}$ up-spin atoms and
$N_{\downarrow}$ down-spin atoms, showing that the number of nodes in the
pairing density is $N_{\uparrow}-N_{\downarrow}$. Therefore, to render a
straightforward pairing topology we focus on the few-fermion system. The
simple spatial distribution couples to the trap oblateness and a central
barrier allowing us to propose a direct experimental probe of the pairing
inhomogeneity. The two up and one down-spin excited states are prototypes
for the ground state of many-body systems. Moreover, with more than eight
atoms trapped the system approaches the macroscopic superfluid state,
motivating our program to probe the infinite-body system from a few-atom
standpoint.

\section{Formalism}

A fermionic gas of two hyperfine states is tightly confined to realize the
Hamiltonian
$\hat{H}=-\hbar^{2}\nabla^{2}/2m+m\omega_{\perp}^{2}(x^{2}+y^{2})/2+
m\omega_{\parallel}^{2}z^{2}/2+g(\vec{r}_{1}-\vec{r}_{2})$, with $m$ the
atomic mass, and a harmonic oscillator length
$a_{\parallel}=\sqrt{\hbar/m\omega_{\parallel}}$. We parameterize the
interspecies potential $g(\vec{r})=-U\Theta(R-|\vec{r}|)$ through an s-wave
scattering length $a=R[1-\tan(R\sqrt{mU}/\hbar)\hbar/R\sqrt{mU}]$. We denote
the general state \nud{N_{\uparrow}}{N_{\downarrow}}, and the (excited)
state with symmetry $\alpha\in\{s,p,d,f,g\}$ as
\eud{N_{\uparrow}}{N_{\downarrow}}{\alpha}.

Our main tool to study the ground and excited states is exact
diagonalization.  We use the Gaussian orbitals of the trapping potential
$\phi_{n_\mathrm{x}n_\mathrm{y}n_\mathrm{z}}(x,y,z)$ as the one-particle
basis functions for the calculation, retaining all orbitals that satisfy
$(n_\mathrm{x},n_\mathrm{y},n_\mathrm{z})\le5$. The matrix elements of the
interaction in this basis set are evaluated numerically.  We construct the
$(N_\uparrow,N_\downarrow)$ Slater determinants in this basis set and retain
the $10,\!000$ determinants with lowest non-interacting energy to form a
many-body basis set in which to construct the Hamiltonian matrix. We
diagonalize the matrix to obtain the energy eigenstates $\{ \psi_m(a)
\}$. We can connect the eigenstates at neighboring values of the scattering
length $a_i$ and $a_{i+1}$ with maximal overlap
$\langle\psi_m(a_i)|\psi_n(a_{i+1})\rangle$ to build up the band structure.

\section{Ground state energy}

\begin{figure}
 \begin{tabular}{ll}
 (a) \eud{2}{1}{\alpha} energy states & (b) Energy differences\\
 \includegraphics[width=0.485\hsize]{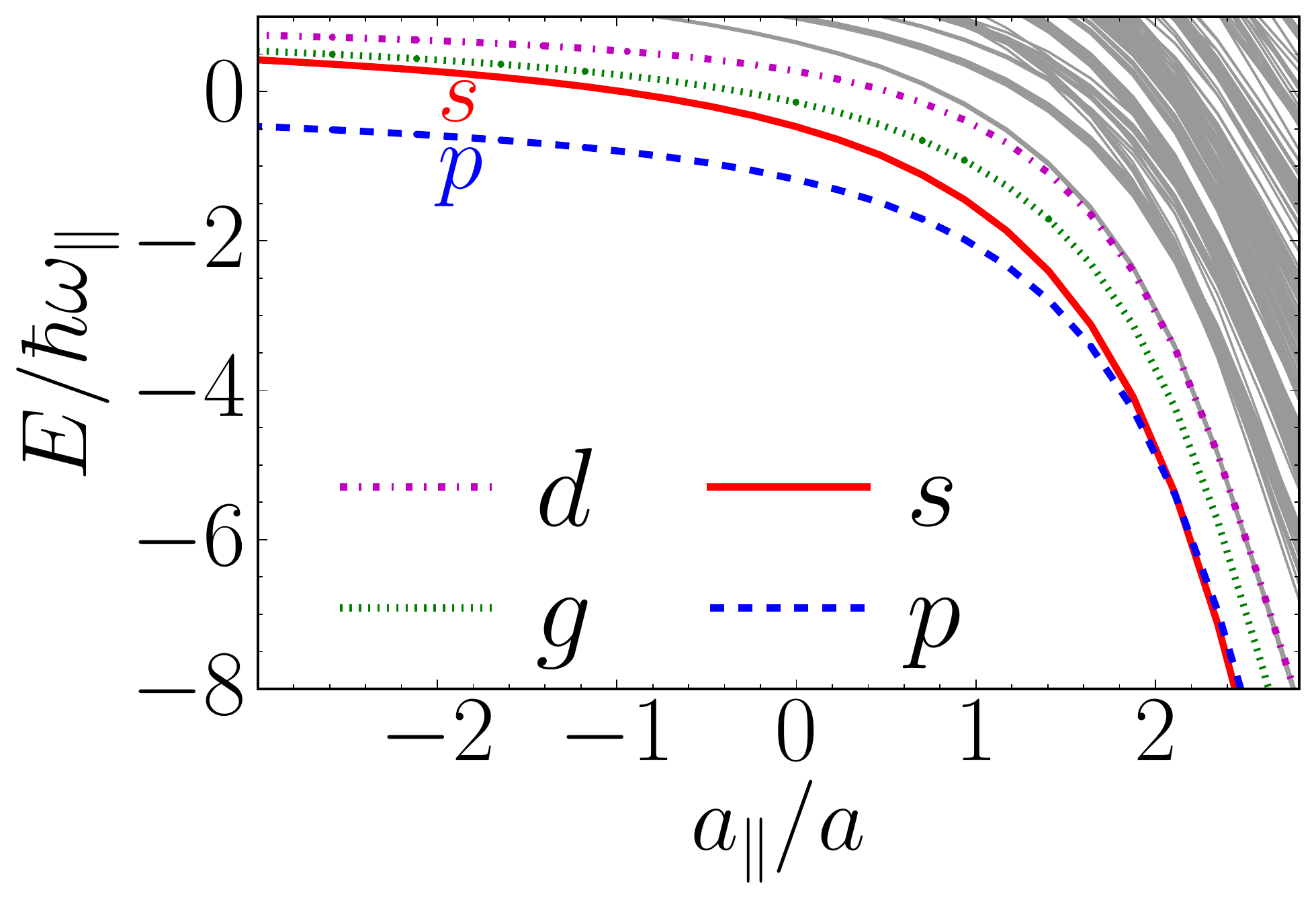}&
 \includegraphics[width=0.485\hsize]{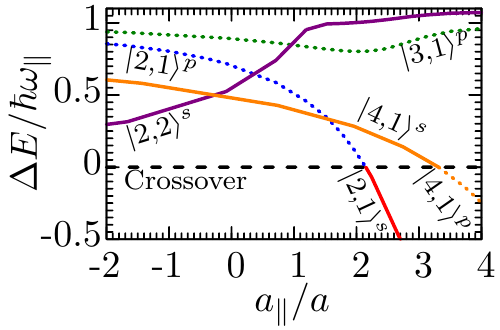}
 \end{tabular}
 \caption{(Color online) %
   (a) The energy bands of the \nud{2}{1} system, highlighting the p-wave wave state
   (blue dashed), s-wave state (red solid), g-wave (green dots), d-wave
   (magenta dot-dash), and higher excited states in gray.
   (b) The difference in energy between the two lowest eigenstates with
   increasing number of atoms. Solid lines denote an s-wave ground state and
   dashed a p-wave ground state.  Different colors denote the labeled
   states, with the same colors depicting the \nud{2}{1} state in both (a)
   and (b).  The horizontal black dashed line delineates the crossover
   between the two lowest eigenstates.}
 \label{fig:Perturbative}
\end{figure}

We first study the ground state energy and symmetry of the atoms in a
spherical trap with $\omega_{\perp}=\omega_{\parallel}$. We start from the
simplest interacting system \nud{1}{1}. At weak interactions both atoms
occupy the $\phi_{000}$ orbital so that the ground state is spherically
symmetric. With increasing interactions the atoms adiabatically evolve into
a tightly bound molecule that retains the same spherical symmetry. The bound
fermionic atoms can now be regarded as a bosonic particle, but to expose
this change in the underlying particle statistics we must introduce a second
up-spin atom.

In the weakly interacting \nud{2}{1} system the new up-spin atom is forced
by Pauli exclusion to enter one of the triply degenerate
($\phi_{100},\phi_{010},\phi_{001}$) orbitals possessing p-wave
symmetry. The next excited state shown in \figref{fig:Perturbative}(a)
places that atom into a singly degenerate superposition of the
$(\phi_{200},\phi_{020},\phi_{002})$ orbitals which will have s-wave
symmetry. Above this there is a nine-fold degenerate g-wave state (the
ground p-wave state excited with center-of-mass motion so that it has triple
the degeneracy of the p-wave state), then a five-fold degenerate d-wave
state with zero center-of-mass motion, and above that many other
disconnected excited states. Previous analysis of this system was performed
in the center-of-mass frame so neglected the g-wave
state~\cite{Stetcu07,Kestner07}.  On entering the strongly attractive regime
an up and down-spin atom bind into a bosonic molecule, removing the Pauli
blocking, leaving the excess up-spin atom in the $\phi_{000}$ orbital. There
must be a crossing from the weakly interacting p-wave to the strongly
interacting s-wave regime. In \figref{fig:Perturbative}(b) we plot the
energy difference between the ground and first excited states. This shows
that the transition from p to s-wave symmetry occurs at
$a=0.46a_{\parallel}$, a crossing phenomenon similar to that suggested in
Refs.~\cite{Stetcu07,Kestner07,Baur10}.  In general the up and down-spin
atoms bind in pairs, leaving any excess majority spins to fill orbitals as
if they were non-interacting, and adopting the symmetry of that state. This
can lead to a changing symmetry of the ground state that betrays the innate
inhomogeneous pairing.

Having seen the consequences of the ground state symmetry changing with the
introduction of the extra up-spin atom we now introduce a further up-spin
atom giving \nud{3}{1}. The weakly interacting ground state carries p-wave
symmetry that adiabatically connects with the strongly interacting ground
state where the two excess up-spin atoms also have p-wave
symmetry. Therefore \figref{fig:Perturbative}(b) shows that this triply
degenerate state is always lower in energy than the next family of excited
states so the system should display inhomogeneous pairing at all interaction
strengths. We next examine the \nud{4}{1} system. In the weakly interacting
regime the majority spins occupy a full shell of
($\phi_{100},\phi_{010},\phi_{001}$) orbitals so the ground state has s-wave
symmetry. In the strongly interacting regime one of the up-spin atoms is
bound to the down spin at $a=0.31a_{\parallel}$, fragmenting the full shell
into a p-wave ground state. Finally we verify that the \nud{2}{2} system has
a spherically symmetric ground state at all interaction strengths indicating
that this balanced system will not display inhomogeneous pairing. The first
excited states display d-wave symmetry. Although they are degenerate with
the ground state in the non-interacting limit, the energy difference grows
rapidly with rising interactions. When the energy difference exceeds
$\hbar\omega_{\parallel}$ a p-wave state (the s-wave state excited with
center-of-mass motion) becomes the new lowest excited state. This crossing
produces the kink at $a\approx1.2a_{\parallel}$.

\section{Inhomogeneous pairing}

\begin{figure}
\begin{tabular}{m{0.315\hsize}m{0.315\hsize}m{0.315\hsize}}
 (a) \nud{2}{1};\nud{3}{2};\nud{4}{3}&
 (b) \nud{3}{1};\nud{4}{2}&(c) \nud{4}{1}\\
 \includegraphics[width=\linewidth]{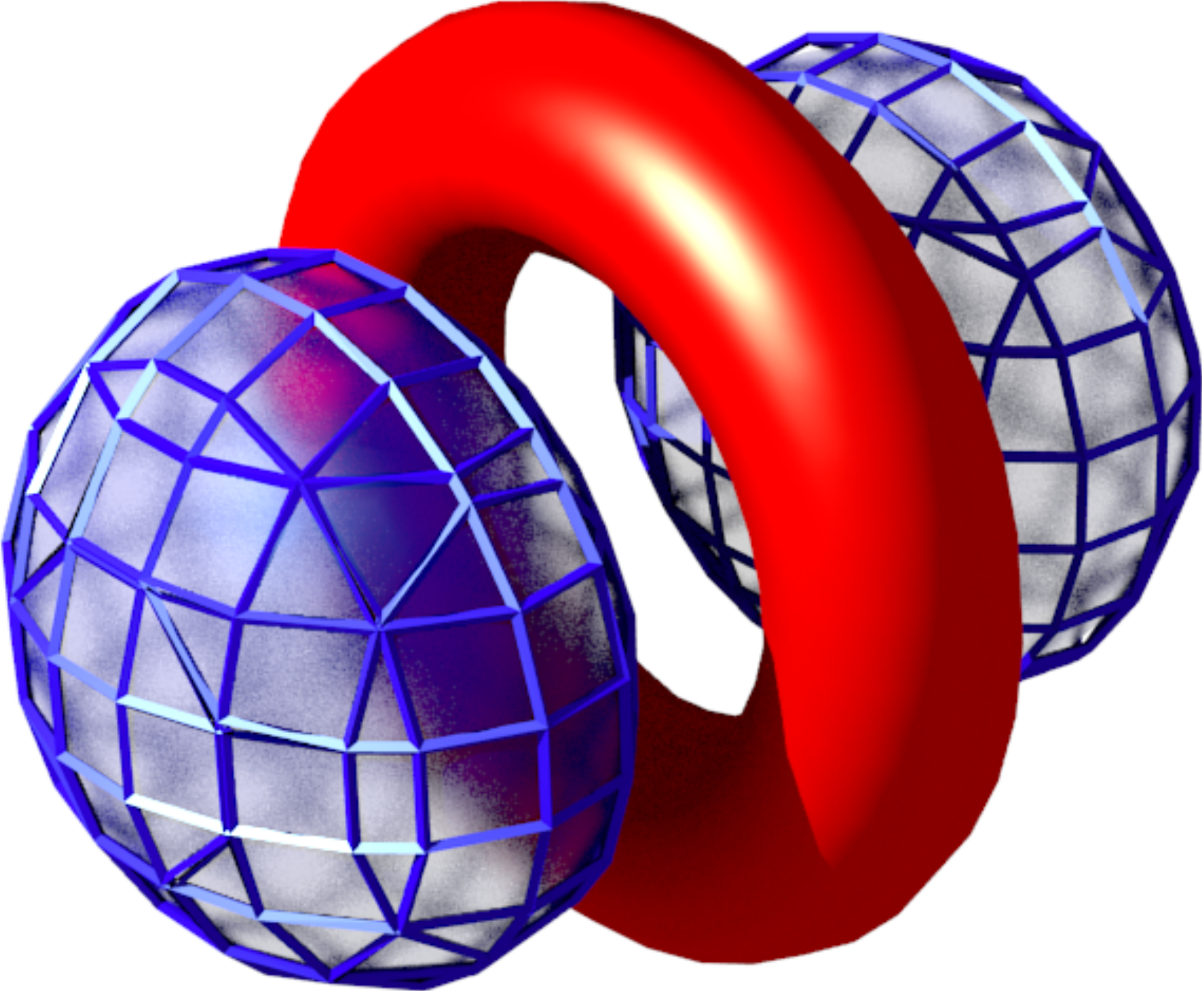}&
 \includegraphics[width=\linewidth]{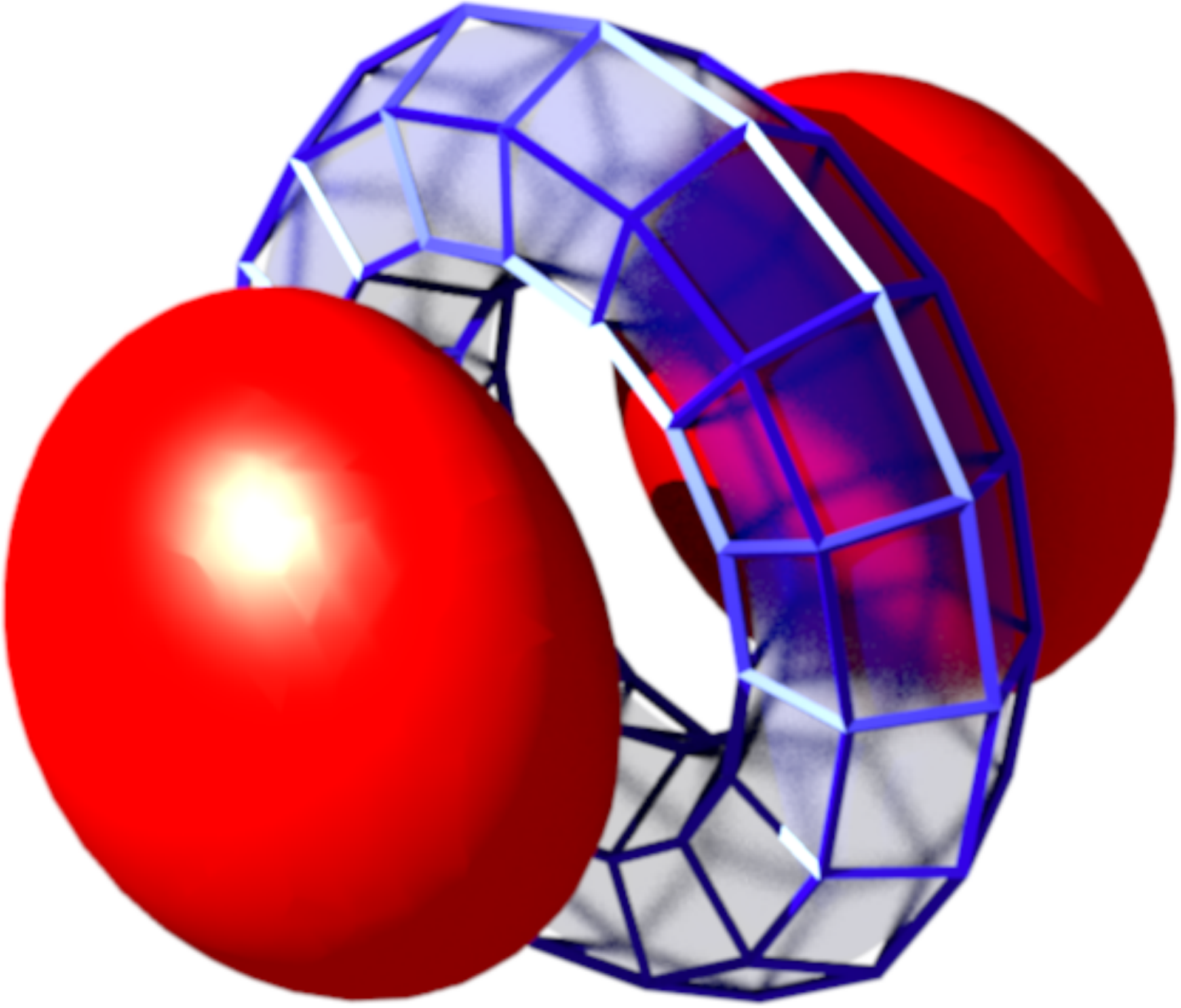}&
 \includegraphics[width=\linewidth]{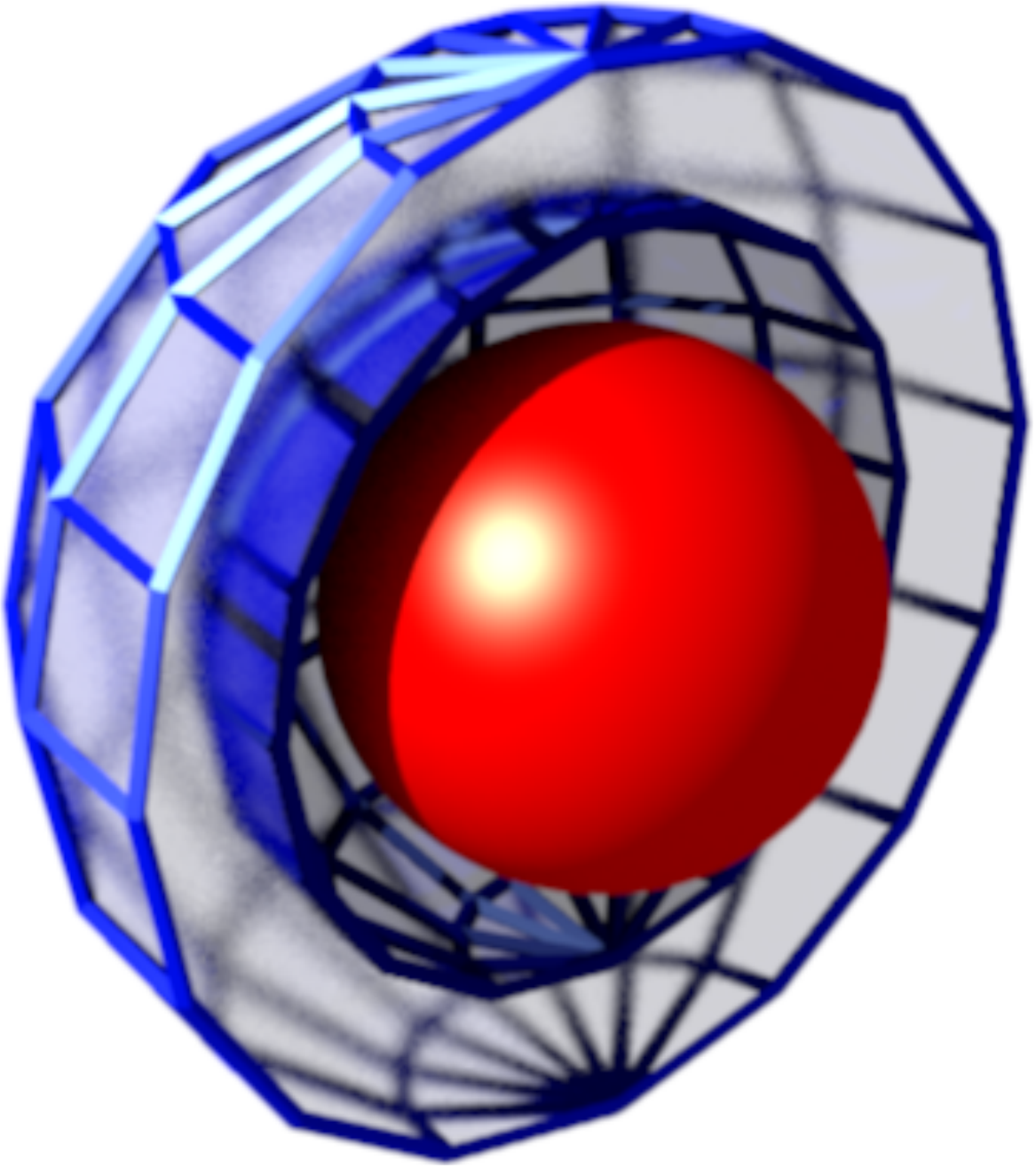}\\
 (d) \eud{2}{1}{s};\nud{5}{1}&
 (e) \eud{2}{1}{d};\nud{5}{2}&
 (f) \eud{2}{1}{g}\\
 \includegraphics[width=\linewidth]{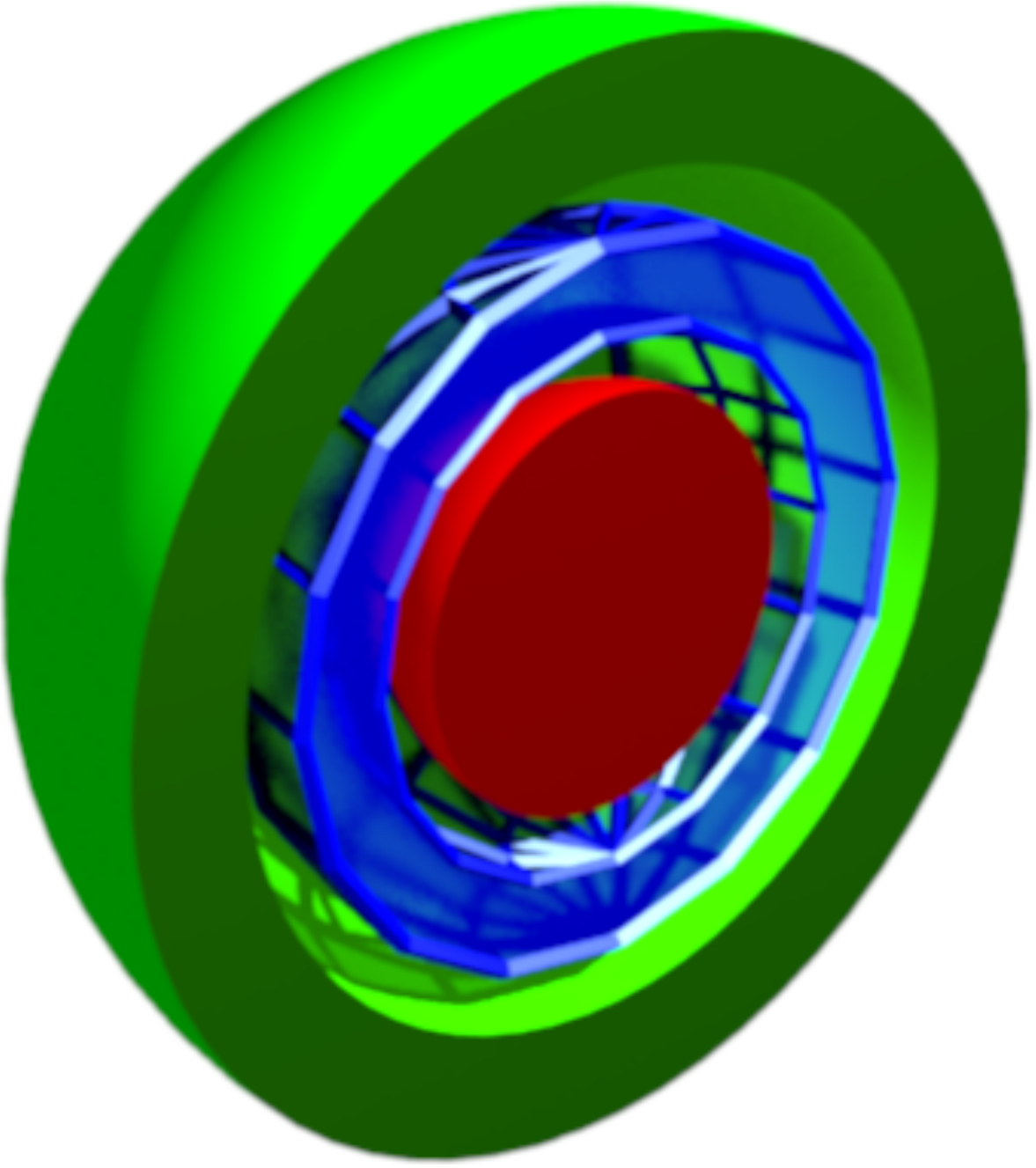}&
 \includegraphics[width=\linewidth]{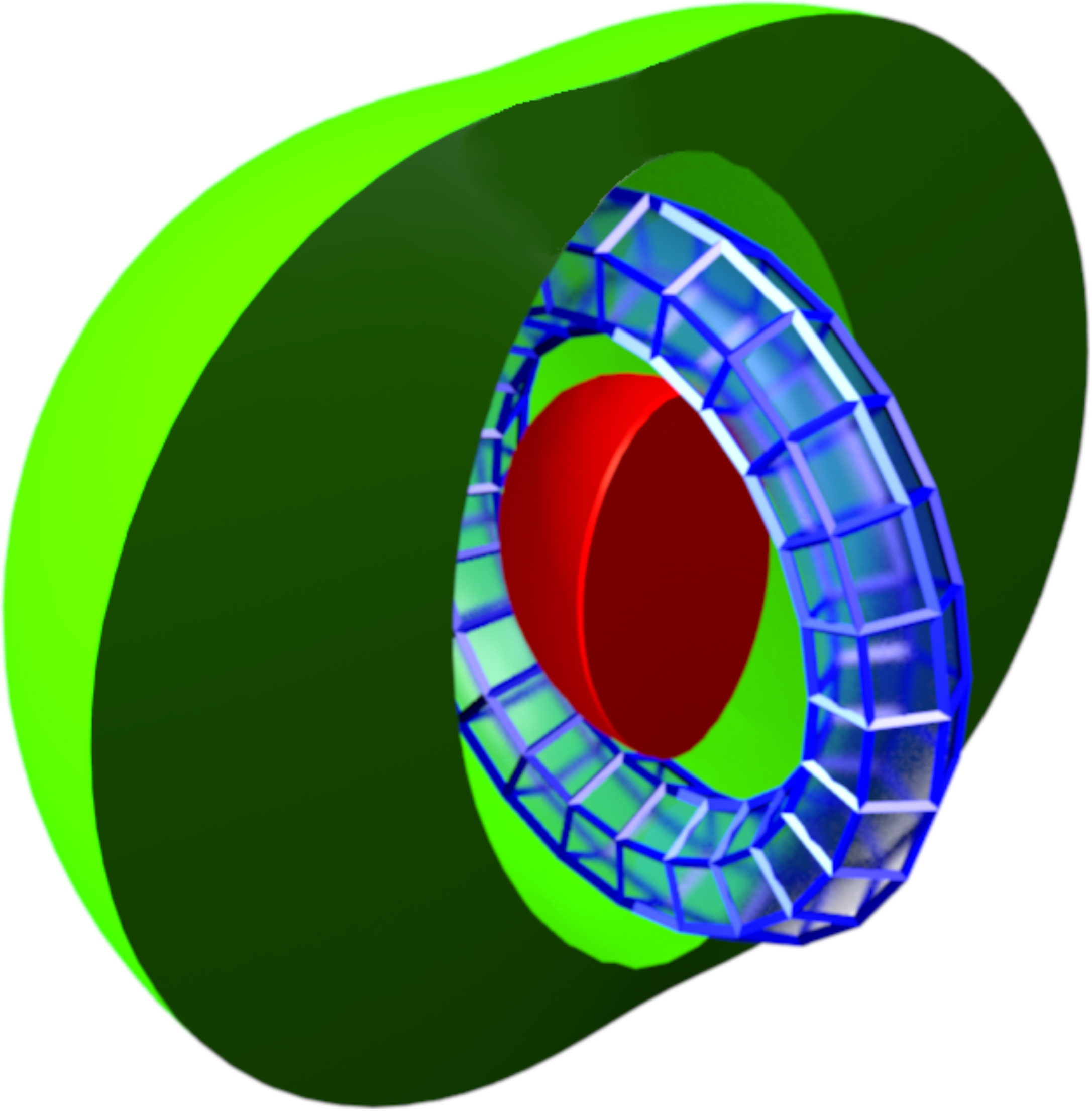}&
 \includegraphics[width=\linewidth]{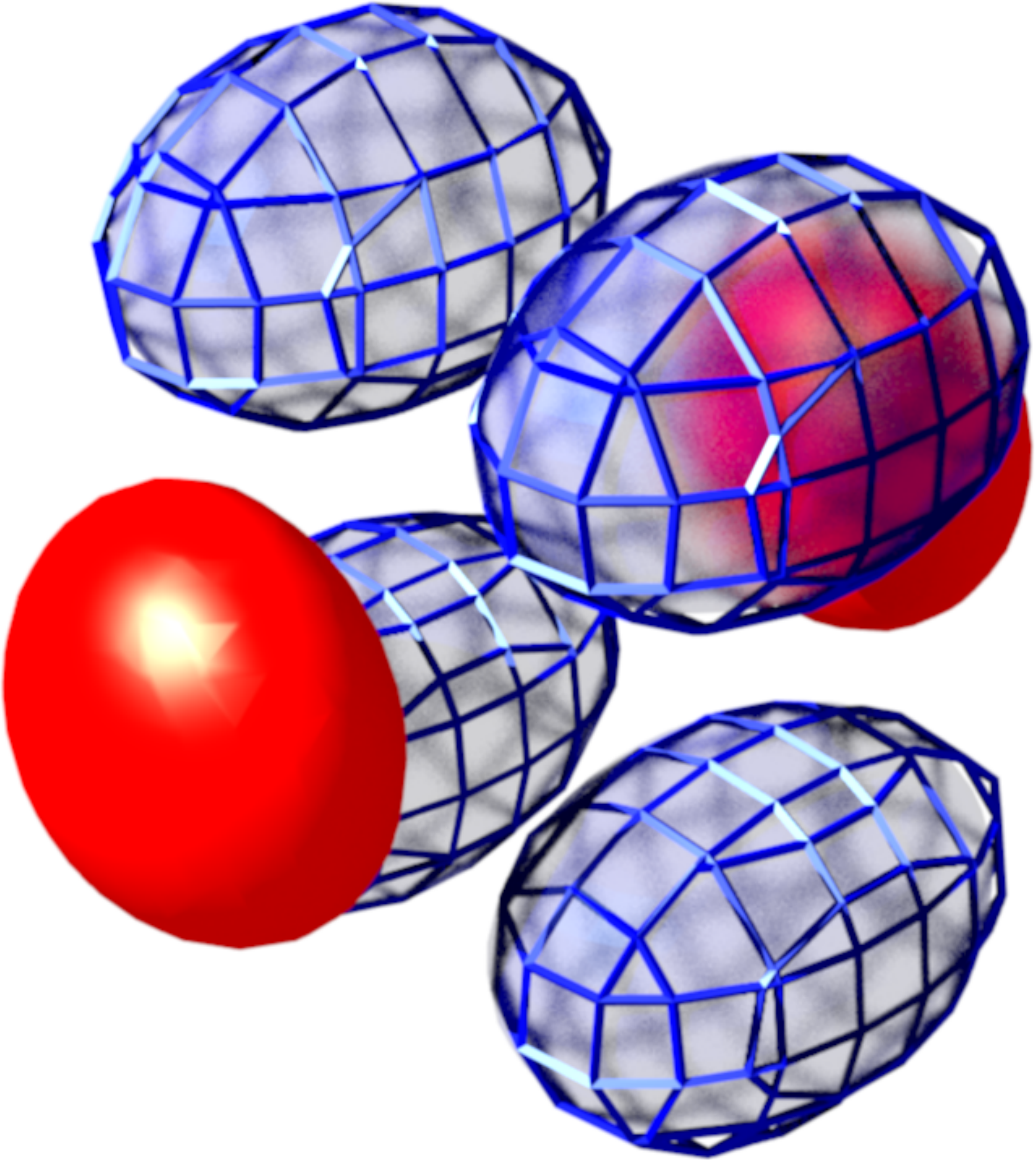}
 \end{tabular}
 \begin{tabular}{ll}
 (g) \nud{2}{1} components&
 (h) Pairing in one dimension\\
 \includegraphics[width=0.485\hsize]{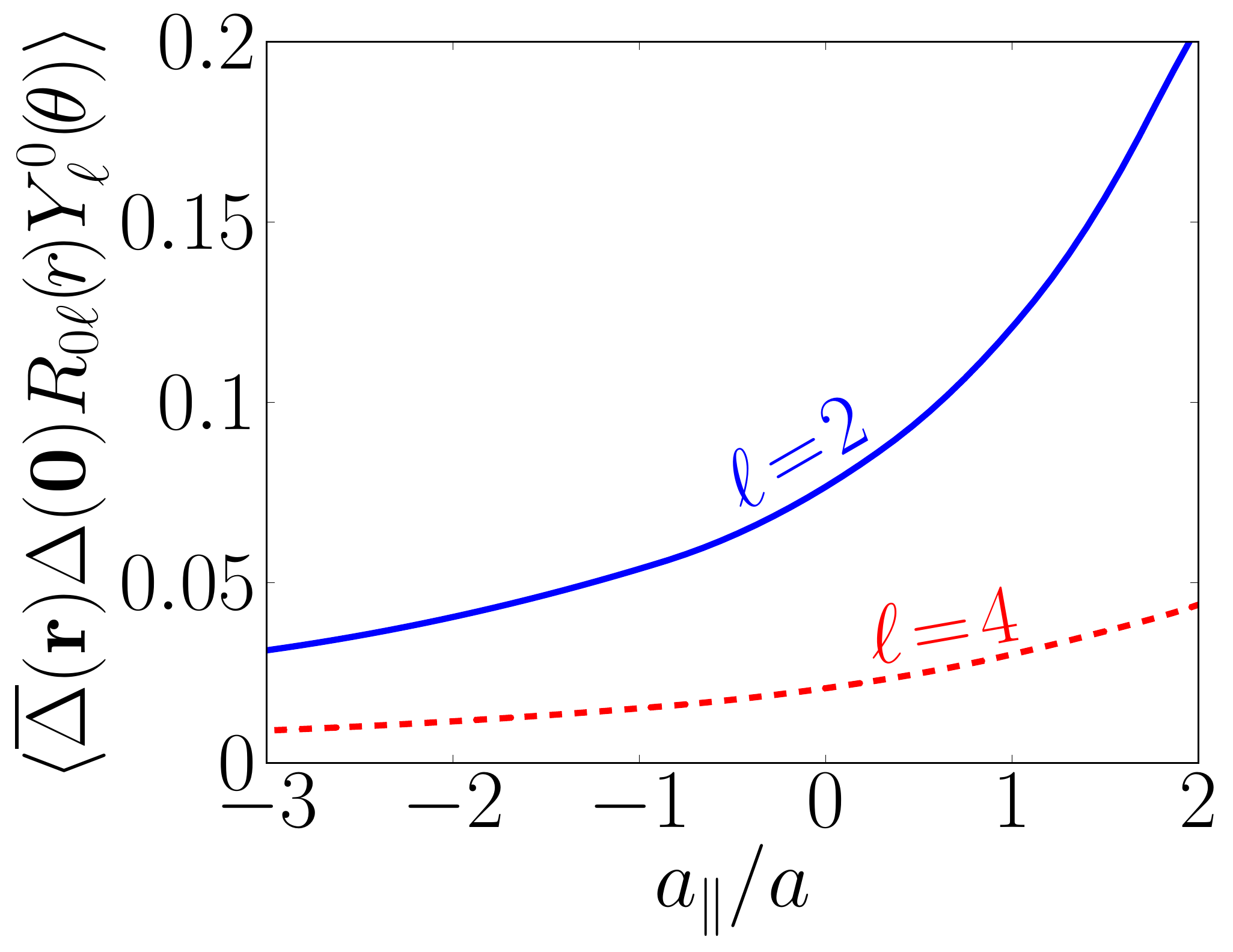}&
 \includegraphics[width=0.485\hsize]{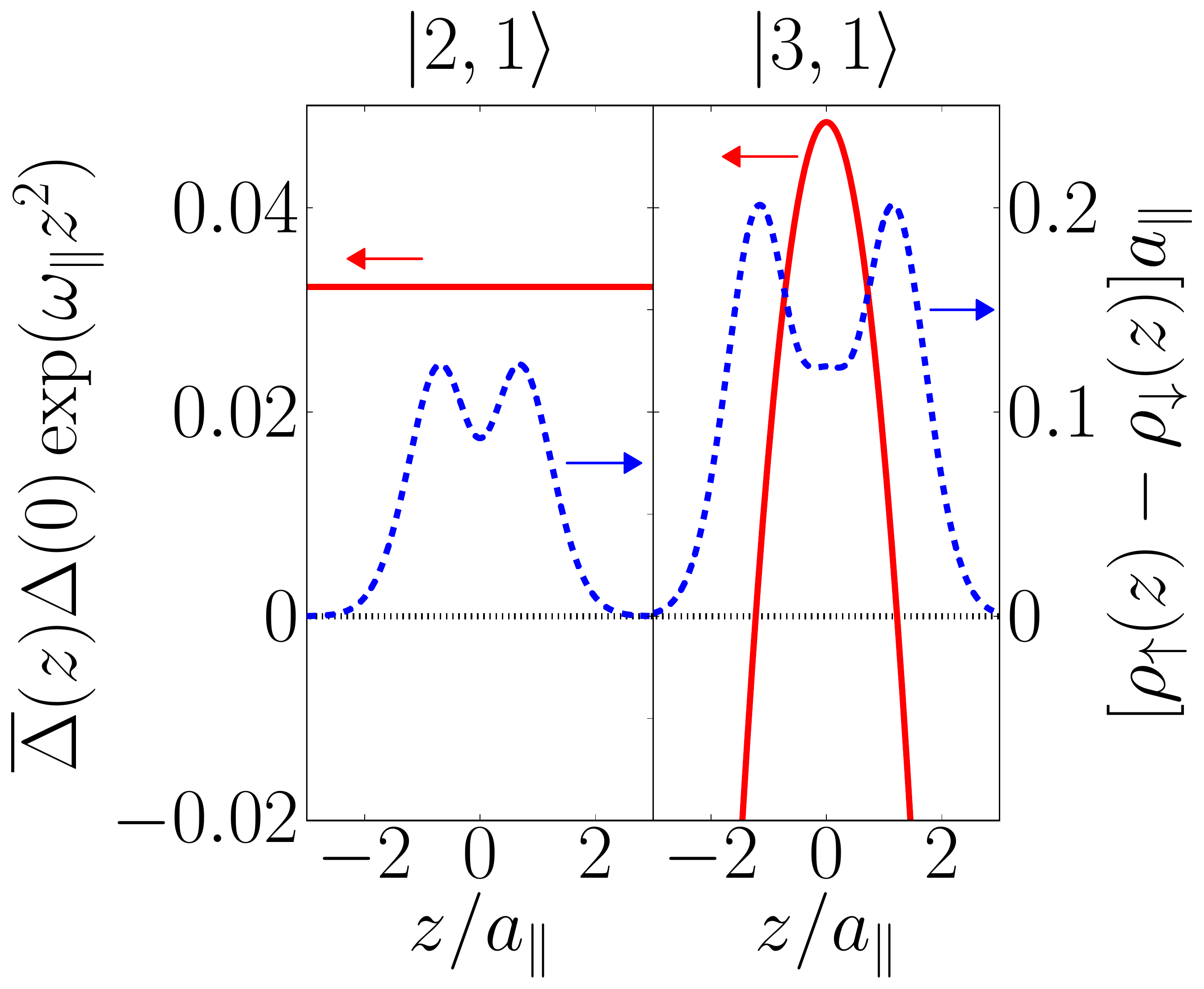}
 \end{tabular}
 \caption{(Color online) (a-f) The isosurfaces of the pairing density
   (positive: red, negative: green), and excess majority spin density (blue
   mesh). (g) the angular spherical harmonic components of the \nud{2}{1}
   system where $R_{n\ell}$ are the radial solutions for principal quantum
   number $n$ and orbital quantum number $\ell$, and $Y_{\ell}^{m}$ denotes
   the spherical harmonic function with projected angular momentum quantum
   number $m$. (h) shows the pairing density (red solid) and excess spin
   density (blue dashed) in a one-dimensional system.}
 \label{fig:CorrFig}
\end{figure}

In the presence of a population imbalance the ground state symmetry can
switch with interaction strength, raising the possibility of inhomogeneous
pairing. This motivates us to study the underlying pairing density. We
measure the pairing correlations with the expectation value
$\bar{\Delta}(\vec{r})\Delta(\vec{0})=\langle
c_{\uparrow}^{\dagger}(\vec{r})c_{\downarrow}^{\dagger}(\vec{r})
c_{\downarrow}(\vec{0})c_{\uparrow}(\vec{0})\rangle$ that explicitly
conserves the number of atoms. In \figref{fig:CorrFig} we compare the
isosurfaces of equal spin imbalance $\langle
c_{\uparrow}^{\dagger}c_{\uparrow}-
c_{\downarrow}^{\dagger}c_{\downarrow}\rangle$ with the interaction strength
independent isosurface of equal pairing growth rate,
$\diffd(\bar{\Delta}\Delta)/\diffd a|_{a=0}$.

We start with \nud{2}{1} atoms, the minimal system that exhibits an
inhomogeneous ground state. Without loss of generality we place the upper
majority spin atom into the $\phi_{001}$ orbital, fixing the excess density
along the z-axis as shown in \figref{fig:CorrFig}(a).  Within first order
perturbation theory the ground state, $\cdu{000}\cdu{001}\cdd{000}\vac$,
couples to the transverse states $\cdu{100}\cdu{001}\cdd{100}\vac$ and
$\cdu{010}\cdu{001}\cdd{010}\vac$. The induced pairing is inhomogeneous as
Pauli blocking prevents the atoms from coupling to
$\cdu{001}\cdu{001}\cdd{001}\vac$. The pairing correlations
$\bar{\Delta}(\vec{r})\Delta(\vec{0})\propto(x^2+y^2)\exp(-\omega_{\parallel}r^2)$
and excess spin density $\sim z^2\exp(-\omega_{\parallel}r^2)$ yield the
isosurfaces shown in \figref{fig:CorrFig}(a). The pairing is peaked in the
regions of low excess up-spin atom density. This distribution is similar to
that proposed for the low-density one-dimensional FFLO
state~\cite{Buzdin83,Machida84}. We can probe the emergence of inhomogeneous
pairing correlations by studying the spherical harmonics present in the
pairing function in \figref{fig:CorrFig}(g). The $\ell=2$ component grows
rapidly with interaction strength signifying the pairing increasing only
within the torus, and the $\ell=4$ component increases so that the pairing
can rise abruptly at the boundary between torus and the excess majority spin
density. The excess majority spin mostly remains in the $\phi_{001}$
orbital.

Now that we have studied the inhomogeneous pairing of the \nud{2}{1} ground
state we turn to re-examine the excited states from
\figref{fig:Perturbative}(a). These are not only accessible through RF
transfers, but moreover act as precursors of states containing more atoms so
will help develop our intuition.  We enumerate the three lowest excited
states in \figref{fig:CorrFig}(d-f) that all exhibit inhomogeneous
pairing. The \eud{2}{1}{s} state in \figref{fig:CorrFig}(d) is formed by
taking the up-spin atom from the $\phi_{001}$ orbital and exciting it into
the spherically symmetric linear combination of the
($\phi_{200}$,$\phi_{020}$,$\phi_{002}$) orbitals. The majority spin atom
lies in a shell at the node of the pairing density. The \eud{2}{1}{d} state
in \figref{fig:CorrFig}(e) is formed by exciting an atom out of the
$\phi_{000}$ orbital and into the $(\phi_{200}+\phi_{020})/2-\phi_{002}$
orbital. The excess spin now lies in a torus that defines the pairing node.
The \eud{2}{1}{g} state \figref{fig:CorrFig}(f) is formed by exciting
\nud{2}{1} with center-of-mass motion in the x-y plane, thereby breaking the
cylindrical symmetry. These excited states contain low energy vacant
orbitals (e.g. $\phi_{000}$) into which we can insert further atoms to form
new ground states.

With our study of the minimal \nud{2}{1} system complete we now study
systems containing more atoms.  Firstly, we examine the \nud{3}{1}
system. \figref{fig:CorrFig}(b) reveals that the majority spins lie in the
x-y plane focusing the pairing along the z-axis. The \nud{4}{1} system has a
full $n=1$ shell of excess spin atoms. The pairing density is spherically
symmetric, enveloped by the excess spin density. The conformation of the
nodal surface is commensurate with the number of excess fermions.  To verify
this conjecture we first look at the \nud{3}{2} and \nud{4}{3}
systems. Although the occupied orbitals contrast to the \nud{2}{1} system,
each has one excess spin fermion with an identical pairing
structure. Likewise the \nud{4}{2} system has an identical topology to the
\nud{3}{1}. The \nud{3}{1};\nud{4}{2};\nud{4}{1} states have no counterpart
in the spectrum of the \nud{2}{1} excitations because they rely on the
occupation of the ($\phi_{100}$,$\phi_{010}$) orbitals that do not couple to
the \nud{2}{1} state so to connect with the excited states we now turn to
systems containing more atoms.

In the \nud{5}{1} system \figref{fig:CorrFig}(d) the new atoms first fill
the ($\phi_{100}$,$\phi_{010}$) orbitals and then the final new atom enters
a linear combination of the ($\phi_{200}$,$\phi_{020}$,$\phi_{002}$)
orbitals, rendering a spherically symmetric state. This can be regarded as
the \eud{2}{1}{s} state but with the
($\phi_{100}$,$\phi_{010}$,$\phi_{001}$) orbitals filled and so adopts the
same topology. Likewise the \nud{5}{2} system can be regarded as the
\eud{2}{1}{d} state with the
($\phi_{000}$,$\phi_{100}$,$\phi_{010}$,$\phi_{001}$) orbitals filled. This
state with 3 excess fermions notably has a different topology to the
\nud{4}{1} state with equal imbalance due to the newly occupied $n=2$ shell
having greater degeneracy with atoms entering the $\phi_{110}$ orbitals that
induce d-wave character. The ability to build many-particle states out of
the excited states of a few-atom system illustrates the utility of studying
the few-atom systems to understand many-body states. In
\figref{fig:PhaseDiagPlot}(b) we connect to the macroscopic FFLO state by
studying how the energy varies with system size, and compare to the 3D
polaron limit that can be solved analytically~\cite{Chevy10}. With more than
eight atoms trapped the energy approaches the infinite-body polaron limit,
showing that the few-atom system directly links to the macroscopic state.

Finally, we examine the pairing in a one-dimensional system with
$\omega_{\perp}=10\omega_{\parallel}$. Starting again with \nud{2}{1} the
pairing is concentrated at the center, forcing the excess majority atom
outwards. This configuration is analogous to the \nud{2}{1} in the spherical
trap that we studied earlier, projected onto the z-axis. With further atoms
present, such as \nud{3}{1} each excess up-spin atom sits at
$z\approx\pm\sqrt{3/2}a_{\parallel}$, with the pairing changing sign as it crosses
through a node at these points. This confirms our picture of the majority
spin atom defining nodes in the pairing correlations, similar to that
proposed for the low-density FFLO state~\cite{Buzdin83,Machida84}.  Though
this one-dimensional system is realizable, the spatial inhomogeneity would
be difficult to probe in experiment. Instead we now return to the
three-dimensional system and exploit the state degeneracy to propose an
experimental observable.

\section{Experimental observation}

\begin{figure}
  \begin{tabular}{ll}
  (a) State diagram & (b) System size\\
  \includegraphics[width=0.485\hsize]{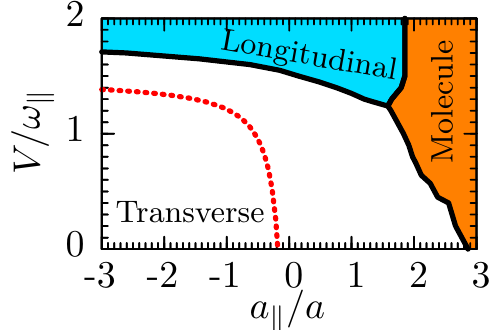}&
  \includegraphics[width=0.485\hsize]{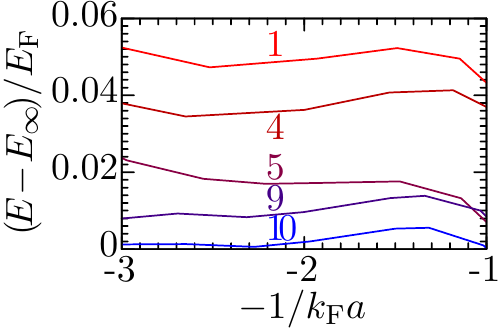}
  \end{tabular}
  \caption{(Color online)
    (a) The ground state for two up and one down-spin atom
    in the trap $\omega_{\perp}=0.5\omega_{\parallel}$ with changing barrier
    height $V$ and interaction strength $a$. The blue shaded region denotes
    the longitudinal ground state, the white area the transverse state, and
    orange the molecular state. The red dashed line denotes the
    longitudinal-transverse boundary predicted by perturbation theory.
    (b) The energy of a single down spin embedded in $N$ up-spin atoms
    compared with the infinite system size limit, with $k_{\text{F}}$
    defined from the non-interacting Fermi energy.}
\label{fig:PhaseDiagPlot}
\end{figure}

Following the emergence of inhomogeneous pairing we now take advantage of
the simple spatial variation of the pairing in the \nud{2}{1} FFLO state to
propose an experimental observable.  Two parameters couple to the angular
pairing oscillations: a pancake trapping potential with
$\omega_{\parallel}>\omega_{\perp}$ encourages the unpaired majority spin
atom into a doubly degenerate transverse orbital ($\phi_{100}$ or
$\phi_{010}$), whereas a central barrier $V\exp(-\omega_{\text{B}}z^2)$
favors occupation of the singly degenerate longitudinal orbital
($\phi_{001}$) that has a node over the barrier. Strong inhomogeneous
pairing attracts density towards the energetically costly central barrier so
to minimize that density the system favors the occupation of the longitudinal
state. This transition provides a direct probe of the inhomogeneous
pairing.

We use both exact diagonalization and first order perturbation theory to
probe the boundary between the transverse and longitudinal
states. Starting from the non-interacting states in the absence of the
central barrier we include both the interactions and the barrier through
first order perturbation theory to yield the estimates for the ground state
energy
\begin{widetext}
\begin{align}
\text{Longitudinal, singly degenerate}\quad&
\frac{5}{2}\omega_{\parallel}+3\omega_{\perp}
+\sqrt{\frac{\omega_{\parallel}}{\omega_{\parallel}+\omega_{\text{B}}}}V\left(2
+\frac{\omega_{\parallel}}{\omega_{\parallel}+\omega_{\text{B}}}\right)
+\frac{a}{a_{\parallel}}\omega_{\perp}\sqrt{\frac{2}{\pi}}
\left(\frac{3}{2}-\frac{4\sqrt{2}}{\pi}\frac{V}{\omega_{\parallel}+\omega_{\text{B}}}\right)\neweqnline
\text{Transverse, doubly degenerate}\quad&\frac{3}{2}\omega_{\parallel}+4\omega_{\perp}
+3\sqrt{\frac{\omega_{\parallel}}{\omega_{\parallel}+\omega_{\text{B}}}}V
+\frac{a}{a_{\parallel}}\omega_{\perp}\sqrt{\frac{2}{\pi}}
\left(\frac{3}{2}-\frac{6\sqrt{2}}{\pi}\frac{V}{\omega_{\parallel}+\omega_{\text{B}}}\right)\punc{.}
\end{align}
\end{widetext}
Setting the two energies equal predicts a crossover at
\begin{align}
 V\!=\!\frac{(\omega_{\text{B}}\!+\!\omega_{\parallel})^{3/2}
 (\omega_{\parallel}\!-\!\omega_{\!\perp})}
 {\omega_{\text{B}}\sqrt{\omega_{\parallel}}}
 \!+\!\frac{4(\omega_{\text{B}}\!+\!\omega_{\parallel})^2
 (\omega_{\parallel}\!-\!\omega_{\!\perp})a}
  {\pi^{3/2}\omega_{\parallel}\omega_{\text{B}}^2a_{\parallel}}\!\punc{.}
\end{align}

In \figref{fig:PhaseDiagPlot}(a) we study the behavior in a trap with
ellipticity $\omega_{\perp}=0.5\omega_{\parallel}$ and
$\omega_{\text{B}}=5\omega_{\parallel}$. In the non-interacting system the
crossover between longitudinal and transverse states predicted by exact
diagonalization is at $V\approx1.87\omega_{\parallel}$ that is in good
agreement with the perturbation theory estimate
$V\approx1.46\omega_{\parallel}$. The critical barrier height falls with
rising interaction strength due to the inhomogeneous pairing pulling density
onto the central barrier and so favoring the longitudinal mode that has a
node over the barrier. With spherically symmetric pairing this transition would
be independent of interaction strength so its gradient exposes the
inhomogeneous pairing. The transition could be exposed by starting from four
trapped atoms and tilting the trap so that one atom escapes, with the other
three atoms entering the ground state. The tunneling rate, proportional to
ground state degeneracy, will be twice as large for the transverse mode than
the longitudinal thus mapping the boundary.

As demonstrated in \figref{fig:Perturbative} the atoms can bind into a
singly degenerate BEC state with s-wave symmetry, this gives the molecular
ground state in \figref{fig:PhaseDiagPlot}(a). At $V=0$ this occurs at
$a_{\parallel}/a=2.86$, which is greater than the $a_{\parallel}/a=2.17$ for
the spherically symmetric system due to the lower energy cost of occupying
the transverse state. The critical interaction strength is a minimum at the
transverse/longitudinal boundary at $V\approx1.2\omega_{\parallel}$ since
those two states are most unstable here. The transverse state is more stable
as the central barrier is reduced, whereas the molecule with density sited
over the barrier becomes less favorable with increasing barrier height than
the longitudinal mode that has a node over the barrier.  Systems containing
more atoms will display analogous phase behavior governed by their
symmetries that were studied earlier.

\section{Discussion}

The few trapped fermions pose a nexus between analytically tractable
few-body physics and intractable many-body systems. We have studied the
imbalanced superfluid whose ground state displays inhomogeneous pairing. The
pairing is analogous to the elusive LO state with the unpaired majority
spins residing along the nodes of the pairing order. The \nud{2}{1} atoms
possess a simple pairing topology that couples to the trap ellipticity and
central barrier that can be probed experimentally.  The ground state of more
atoms can be understood in terms of the excited states of the \nud{2}{1}
system. Moreover, the system approaches the infinite-body limit when more
than eight atoms are trapped, so our study has broad implications for the
FFLO phase.

\acknowledgments{The authors thank Gerhard Z\"urn, Thomas Lompe, Selim
  Jochim, Jonathan Lloyd-Williams \& Stefan Baur for useful discussions. POB
  acknowledges the financial support of the EPSRC, and GJC the support of
  Gonville \& Caius College.}

\end{document}